\documentclass{agujournal2019}
\usepackage{color}
\usepackage{url} 
\usepackage{lineno}

\draftfalse

\newcommand{\ssr}{    {Space Sci. Rev.}}

\newcommand{\apjl}{    {Astrophys. J. Lett.}}
\newcommand{\nat}{    {Nature}}
\newcommand{\grl}{    {Geophys. Res. Lett.}}
\newcommand{\jgr}{    {J. Geophys. Res.}}

\usepackage{amssymb}
\usepackage{amsmath, mathtools}

\journalname{\jgr}

\begin{document}

%
%



%
%




\title{Tens to hundreds of keV electron precipitation driven by kinetic Alfv{\'e}n waves during an electron injection}

\authors{Yangyang Shen\affil{1}, Anton V Artemyev\affil{1,2}, Xiao-Jia Zhang\affil{1}, Vassilis Angelopoulos\affil{1}, Ivan Vasko\affil{2,3}, Drew Turner\affil{4}, Ethan Tsai\affil{1}, Colin Wilkins\affil{1}, James M Weygand\affil{1}, Christopher T Russell \affil{1}, Robert E Ergun\affil{5}, Barbara L Giles\affil{6}}

\affiliation{1}{Department of Earth, Planetary, and Space Sciences, University of California, Los Angeles, California, USA}
\affiliation{2}{Space Research Institute of Russian Academy of Sciences, Russia, Moscow}
\affiliation{3}{Space Science Laboratory, University of California, Berkeley, California, USA}
\affiliation{4}{Johns Hopkins University Applied Physics Laboratory, Maryland, USA}
\affiliation{5}{Laboratory for Atmospheric and Space Physics, University of Colorado Boulder, Boulder, CO, USA.}
\affiliation{6}{NASA Goddard Space Flight Center, Greenbelt, MD, USA.}

\correspondingauthor{Yangyang Shen}{yshen@epss.ucla.edu}
\begin{keypoints}
\item Conjugate ELFIN and MMS observations reveal evidence of tens to hundreds of keV electron precipitation associated with kinetic Alfv{\'e}n waves
\item Dipolarized magnetic field gradients and perpendicular magnetic drifts allow Landau resonance between waves and injected electrons 
\item Adiabatic transport and spatially-varying $E\times B$ and grad-$B$ drifts induce perpendicular momentum scattering and electron losses
\end{keypoints}

\begin{abstract}
Electron injections are critical processes associated with magnetospheric substorms, which deposit significant electron energy into the ionosphere. Although wave scattering of $<10$ keV electrons during injections has been well studied, the link between magnetotail electron injections and energetic ($\geq 100$ keV) electron precipitation remains elusive. Using conjugate observations between the ELFIN and Magnetospheric Multiscale (MMS) missions, we present evidence of tens to hundreds of keV electron precipitation to the ionosphere potentially driven by kinetic Alfv{\'e}n waves (KAWs) associated with magnetotail electron injections and magnetic field gradients. Test particle simulations adapted to observations show that dipolarization-front magnetic field gradients and associated $\nabla B$ drifts allow Doppler-shifted Landau resonances between the injected electrons and KAWs, producing electron spatial scattering across the front which results in pitch-angle decreases and subsequent precipitation. Test particle results show that such KAW-driven precipitation can account for ELFIN observations below $\sim$300 keV. 
     
\end{abstract}

\section*{Plain Language Summary}
Energetic electron precipitation from magnetospheric injections has a major impact on magnetosphere-ionosphere coupling. This energy deposition is largely in the form of electron precipitation driven by wave-particle interactions in the magnetotail. Although wave-driven precipitation with energies less than approximately 10 keV has been studied extensively, the link between energetic electron precipitation ($\geq$ $\sim$100 keV) and electron injections remains elusive. Combining observations and simulations, this paper provides evidence of such precipitation driven by kinetic Alfv{\'e}n waves (KAWs), which have been previously observed to be ubiquitously associated with magnetospheric electron injections but have not been considered as an important driver for precipitation of such electrons.

\section{Introduction}

Magnetospheric plasma sheet electron earthward injections feature abrupt and intense flux increases of electrons with energies of tens to hundreds of keV on the nightside magnetotail-inner magnetosphere interface, which are an inherent phenomenon associated with magnetospheric substorms \cite{Akasofu64,McIlwain74,Birn14,Gabrielse14,Turner16}. These energetic electron injections provide a seed population for the radiation belts \cite{Jaynes15:seedelectrons,Turner15} and generate significant magnetospheric electron precipitation to the ionosphere \cite{Clilverd08,Ni16:ssr}.

Electron acceleration in injections often leads to precipitation via wave-particle interactions, so there is an upper limit of the trapped electron fluxes, i.e., the Kennel-Petschek limit \cite{Kennel&Petschek66}. Injected ion and electron distributions can be unstable to various plasma waves, such as whistler-mode chorus waves, electron cyclotron harmonic (ECH) waves, electromagnetic ion cyclotron (EMIC) waves, and nonlinear time domain structures (TDS), which produce diffuse auroras \cite{Thorne10:Nature,Ni16:ssr,Kasahara18:nature,Vasko17:diffusion,Vasko17:diffusion,Shen21:EH} and the loss/acceleration of radiation belt particles \cite{Albert03,Millan&Thorne07,Shprits08:JASTP_local,Li&Hudson19,Thorne21}. Statistical studies have revealed a high correlation between energetic ($>\sim$30 keV) electron injections and ground-based riometer absorption (cosmic radio noise), both at geosynchronous magnetic footprints \cite{Arnoldy&Chan69,Baker81,Spanswick07,Kellerman15} and at the stretched-to-dipolar field transient region up to $L\sim$12 \cite{Clilverd08,Clilverd12,Gabrielse19}. This riometer signal correspondence has been mainly attributed to strong pitch-angle scattering and precipitation of energetic injection electrons \cite{Baker81,Spanswick07}. However, the scattering mechanisms or wave modes driving the energetic precipitation of such injected electron from the plasma sheet to the ionosphere remain elusive. 

Electron scattering by whistler-mode chorus waves has been known as a driver of energetic precipitation from the inner magnetosphere ($L\leq$7) \cite{Horne&Thorne03,Omura&Summers06,Lam10}, but its efficiency of producing strong scattering of injection electrons in the plasma sheet has been questioned in a recent statistical study of 733 dispersionless injections \cite{Ghaffari21:whistler}. Indeed, statistical whistler observations have shown that the occurrence rate and intensity of waves drop significantly beyond $L\sim$8 \cite{Li09:grl,Meredith21}, and resonant field-aligned electron energies hardly reach $\sim$100 keV for parallel-propagating whistlers associated with plasma sheet injections \cite{Li11}. For similar reasons, diffuse auroral precipitation ($<\sim$50 keV) from the outer magnetosphere has been mainly attributed to ECH waves instead of whistlers \cite{Zhang15:ECH,Ni16:ssr}. EMIC waves mainly scatter relativistic ($\sim$MeV) electrons from the dusk and dayside sectors and thus are less likely responsible for nightside plasma sheet electron precipitation \cite{Albert03,Thorne10:GRL,Allen15}. Another precipitation mechanism concerns magnetic field-line curvature scattering, which produces efficient plasma sheet electron pitch-angle isotropization if the magnetic field configuration provides $R_c/\rho_e\leq$8, where $R_c$ is the field line curvature radius and $\rho_e$ is the energetic electron gyroradius \cite{Buchner&Zelenyi89,Sergeev83}. During dipolarizations associated with injections $R_c$ is significantly increased, so curvature scattering will be reduced for injection electrons. The dip in $B_z$ ahead of the dipolarization front can produce transient, localized, and isotropic precipitation \cite{Eshetu18}, but such localized dips cannot be responsible for massive precipitation. MHD ULF waves associated with injections \cite{Shiokawa97,Runov14} have a limited effect on modifying magnetic field gradient scale length $({|\nabla B|}/B)^{-1}$ ($\sim R_E$) and seldom directly impact energetic electrons in pitch angle \cite{Falthammar65,Ukhorskiy&Sitnov13:ssr}. Instead, ULF waves are more likely to contribute by coupling with other kinetic-scale waves \cite{Zhang19:jgr:modulation}, including the generation of kinetic Alfv{\'e}n waves (KAWs) through mode-coupling \cite{Hasegawa&Chen75,Lin12} or phase-mixing \cite{Allan&Wright00} at plasma boundaries, such as dipolarization fronts where injections are seen.

KAWs carrying significant Poynting fluxes have been suggested to be an important pathway of energy transport associated with injections carried by bursty bulk flows (BBFs) in the flow-braking region \cite{Angelopoulos02,Chaston12:fast_flow}. Large-amplitude KAWs have been found to be pervasive within the braking BBFs and injection (dipolarization) fronts \cite{Ergun15}. A high correlation between KAWs and injections has also been reported by \cite{Malaspina15} in the inner magnetosphere, which has further shown that KAW broadband emissions were colocated and comoving with the injection boundary. The movement of injection boundaries is reflected in riometer absorption on the ground, which usually rises and extends westward following the buildup of sustained injection and dipolarization \cite{Gabrielse19}. Transient riometer rises are also associated with auroral streamers, which have been viewed as the signature of BBF channels in the plasma sheet \cite{Henderson98,Lyons12}. Recent global hybrid simulations have provided a comprehensive picture of BBF and KAW generation and propagation within the flow-braking region \cite{Cheng21}. These studies have provided strong evidence that KAWs are correlated with injections, dipolarizations, and braking ion flows, and thus may play a role in energetic precipitation therefrom.    

Conventionally, KAWs are not expected to resonate with injection energetic electrons directly. KAWs have perpendicular wavelengths comparable to the ion thermal gyroradius, and the finite Larmor radius effect produces charge separation and coupling to electrostatic (ion-acoustic) mode, so that a significant parallel-to-{\bf B} electric field develops to maintain charge neutrality and counteracts the electron thermal pressure \cite{Hasegawa76,Lysak&Lotko96}. KAWs parallel electric fields allow electron Landau resonance but typically require the electron velocity to approach the Alfv{\'e}n speed $v_{||}\approx \omega/k_{||}\sim v_A$, limiting the resonant energies to below a few keV \cite{Kletzing94,Watt&Rankin09:prl,Watt&Rankin12} including resonance broadening effects \cite{Artemyev15:jgr:KAW,Damiano15}. One exception is nonlinear stationary inertial Alfv{\'e}n waves, which accelerate counter-propagating electrons greatly exceeding $v_A$ but only apply to the ionospheric low-$\beta$ current sheets with normal plasma drifts \cite{Knudsen96,Liang19:aurora}. Although standing KAWs of field line resonances can pitch-angle scatter relativistic electrons above a few hundred keV through drift-bounce resonance \cite{Chaston18:scattering}, the time scales of such scattering ($\sim$hours) do not allow this mechanism to operate within an injection time period (up to tens of minutes). 

However, when observed by electrons drifting across the local magnetic field with velocity $v_{drift}$, KAW plasma frame $\omega$ can be significantly increased due to Doppler shift and the resonant energy can be shifted to a higher value $v_{||}\sim (\omega-{\bf k_\perp v_{drift}})/k_{||}$. This is possible because KAW perpendicular phase velocity $\omega/k_{\perp}$ can be much less than $v_A$ with $k_{\perp}\gg k_{||}$ \cite{Chaston12:fast_flow}. Coupling between KAW electric fields and particle perpendicular magnetic drifts has been theoretically analyzed by \cite{Johnson&Cheng97} to explain plasma transport at the magnetopause. For electron injections in the tail, equatorial magnetic field gradients associated with dipolarizations \cite{Liu13:DF} provide electron magnetic drifts (e.g., $>$100 km/s for $>$50 keV electrons) comparable to the KAW perpendicular phase speed, potentially moving energetic electrons into Landau resonance with KAWs. Will KAWs drive energetic electron precipitation associated with the plasma sheet injection via this Doppler-shifted Landau resonance?    

In this paper, we present evidence of tens to hundreds of keV electron precipitation driven by KAWs during a magnetotail electron injection, based on observations from the Electron Loss and Fields Investigation (ELFIN) \cite{Angelopoulos20:elfin} and Magnetopsheric Multiscale (MMS) \cite{Burch16} spacecraft. We show results from test particle simulations that demonstrate agreement of the proposed mechanism with such observations.

\section{Data}

We present conjugate observations of a magnetotail electron injection and electron precipitation based on data recorded by MMS and ELFIN on 29 September 2020. We will use the following datasets from MMS:(i) the Fast Plasma Instrument (FPI), which provides electron fluxes within energies of 10 eV--30 keV every $\sim$4.5 s in fast mode \cite{Pollock16:mms}; (ii) the Fly's Eye Energetic Electron Proton Spectrometer (FEEPS), which measures electron fluxes and pitch-angle distributions within energies of 25--650 keV every $\sim$20 s in the spin resolution \cite{Blake16}; (iii) the FIELDS instrument suite \cite{Torbert16:mms,Ergun16:ssr,Lindqvist16,LeContel16:mms,Russell16:mms}, which in fast mode measures DC vector magnetic fields and electric fields at 16 samples per second (sps) and 32 sps, along with wave spectra in frequencies of up to 8 kHz every $\sim$2 s in low-frequency (LF) mode. 

We use data from the ELFIN energetic particle detector for electrons (EPDE) that measures electron fluxes and pitch-angle distributions in the energy range of 50 keV to 5 MeV \cite{Angelopoulos20:elfin}. The ELFIN twin CubeSat (ELFIN-A and ELFIN-B) were launched on 15 September 2018 into polar circular orbits at $\sim$450 km altitude. Mounted on a spinning spacecraft, EPDE has an angular resolution (FWHM) $\sim$22.5$^\circ$ and rotates across an angle of $\sim$24$^\circ$ in $\sim0.18$ s, nominally allowing full pitch angle resolution twice per spin (16 angular sectors in a $\sim$3-s spin period) when the B-field is within $\pm$15$^\circ$ with respect to the spacecraft spin plane (as in our case). Given that the local loss cone is approximately 65$^\circ$ at $\sim$450 km altitude in our event, ELFIN can reliably resolve precipitating, backscattered, and trapped fluxes by averaging measurements from angular sectors within and outside the loss cone. Along-track separation of the identical ELFIN spacecraft has the advantage of resolving the spatio-temporal ambiguity of electron precipitation on time scales of seconds to minutes. 

In addition to spacecraft observations, we also use the horizontal magnetic perturbations, i.e., the northward ($dBn$) and eastward ($dBe$) component from the ground-based magnetometer measurements at Rankin Inlet ($lat\sim$62.82$^\circ$, $lon\sim$267.89$^\circ$) in conjunction with ELFIN measurements. These data are obtained from SuperMAG in 1 sps \cite{Gjerloev12}. Furthermore, we also use a well-developed and validated magnetometer data product of 2D ionospheric currents, applying the spherical elementary current system (SECS) method \cite{Amm&Viljanen99,Weygand11} to a dense network of North-American and Greenland ground-based magnetometer arrays \cite{Mann08,Engebretson95,Russell08:ssr}. With a temporal resolution of 10 s and spatial resolution on the order of $\sim$350 km, dynamic maps of equivalent ionospheric currents (EICs, horizontal currents)
and current amplitudes (SECAs, a proxy for field-aligned currents) allow identification of large-scale substorm current wedges as well as small-scale transient currents associated with injection and dipolarization/BBFs \cite{Panov16}. This identification can help us to pinpoint ELFIN precipitation locations relevant to the magnetospheric injection, which further helps to establish a better conjunction between ELFIN and MMS. 

\section{Observations}
\begin{figure}
\centering
\vspace*{-0.5cm}\hspace*{-1.0cm}\includegraphics[width=1.1\textwidth]{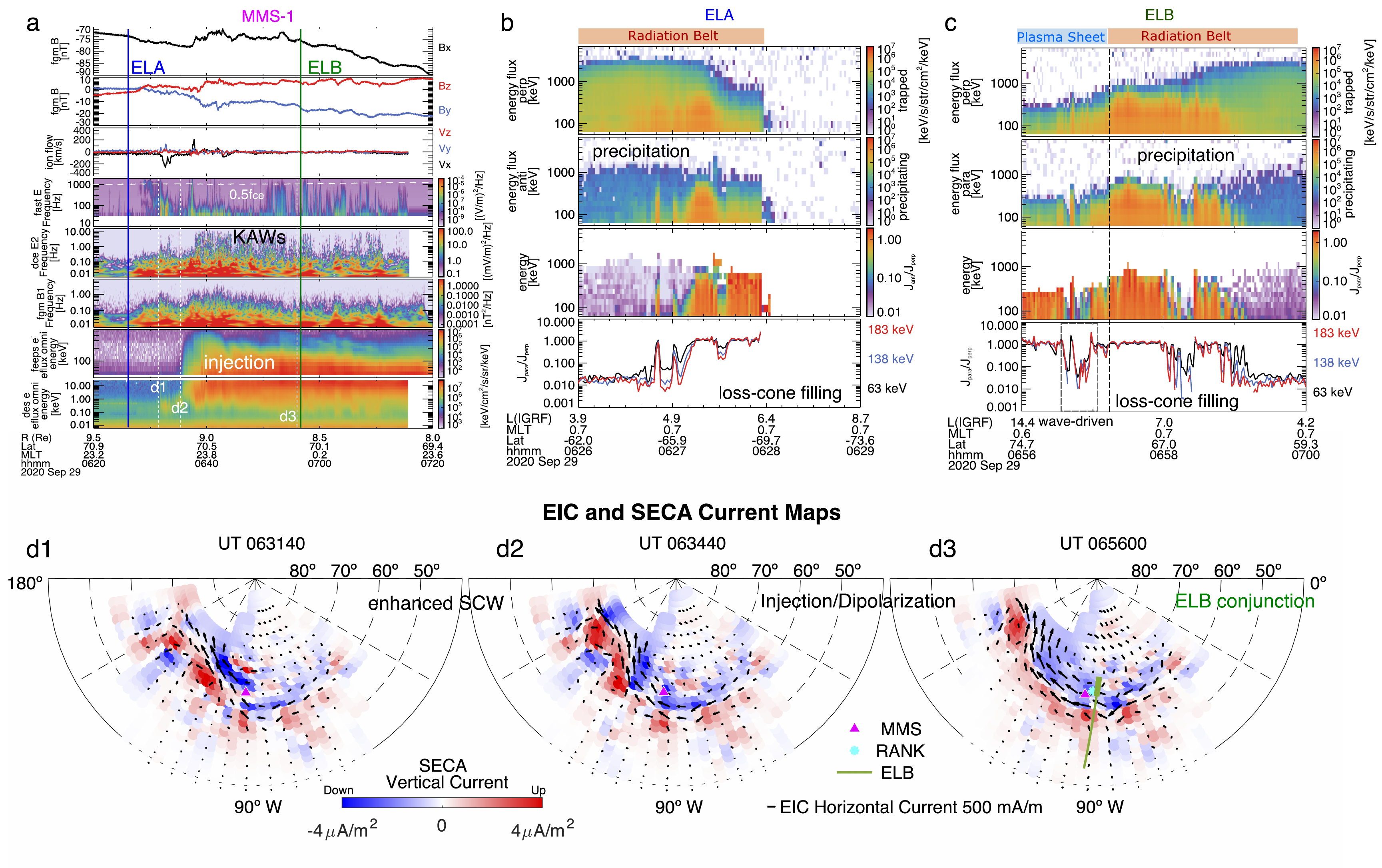}
\caption{MMS-ELFIN conjunction event on 29 September 2020. (a) Panels from top to bottom: MMS $B_x$ in GSM; MMS $B_y$ and $B_z$; MMS ion flows; fast-mode wave electric field spectrogram; DC-coupled perpendicular electric field ($E_2$ component, see the main text) wavelet spectrogram in the frequencies up to 16 Hz; DC-coupled perpendicular magnetic field ($B_1$ component) wavelet spectrogram in the frequencies up to 8 Hz; FEEPS electron energy flux spectra; FPI electron energy flux spectra; the time stamps of the twin ELFIN passes and the ground-inferred current maps (at 100 km altitude) are also shown. (b) ELA trapped and precipitating electron energy fluxes, along with the spectrogram and line plots (63, 138, and 183 keV) of the loss-cone filling ratios. (c) ELB observations during the injection, in the same format as b. Refer to the text for the definition of plasma sheet precipitation. (d) Three snapshots of equivalent ionospheric horizontal currents (EICs) and spherical elementary current amplitudes (SECAs, vertical currents) inferred from magneotometer arrays in geographic coordinates \cite{Weygand11}. The dots for EICs show the location at which the current was determined. The arrow and length of the segment indicate the direction and magnitude. The scale for downward (blue) and upward (red) currents is shown in colorbar. MMS (magenta triangle) and ELB (green line) footprints (TS04 mapping) are shown in the context of the substorm current wedge. ELB plasma sheet precipitation is indicated as a thickened green line. The magnetometer stations of Rankin Inlet (RANK) is shown as the cyan star}. 
\vspace*{-1.3cm}
\label{fig1}
\end{figure}

Fig.~\ref{fig1} presents the plasma sheet electron injection and precipitation event observed from MMS-1, ELFIN-A (ELA), and ELFIN-B (ELB) on 29 September 2020. Fig.~\ref{fig1}a demonstrates MMS-1 observations of the background magnetic field, ion bulk flows, spectra of wave electric and magnetic fields, and spectra of electron energy fluxes at $L\sim$9$R_E$ in the midnight magnetotail. MMS-1 was in the lobe before a sudden, strong electron injection with energies of 100 eV up to 500 keV engulfed the spacecraft near 06:35 UT, after which the injected electrons persisted over 40 min. Accompanying the injection front was a magnetic field $B_z$ increase, signifying the magnetic field dipolarization. More interestingly, enduring broadband waves from sub-Hz up to $\sim$1 kHz were associated with the injected electrons during the entire period. These broadband waves are electromagnetic below a few Hz and become increasingly electrostatic above a few Hz, which are potentially comprised of kinetic Alfv{\'e}n waves and nonlinear time domain structures \cite{Chaston15:kaw,Mozer15}. 

These KAWs can be identified from the fifth and sixth panels of Fig.~\ref{fig1}a, which display the DC-coupled perpendicular electric and magnetic field spectrograms of the low-frequency broadband emission below approximately 10 Hz. We have transformed the measured fields from the Geocentric Solar Magnetospheric (GSM) coordinates into the field-aligned coordinates. The background magnetic field vector is determined by averaging DC-coupled magnetic fields for 3 min. The $E_3$ and $B_3$ components denote field-aligned variations. The $E_1$ and $B_1$ components are perpendicular to $B_3$ and lie in a plane defined by $B_3$ and the geocentric radius vector (see, e.g., \citeA{Rae05}). The $E_2$ and $B_2$ components complete the right-handed orthogonal set.

Before the injection took place, ELA crossed the ionospheric footprint of MMS in the southern hemisphere during 06:24--06:30 UT around magnetic midnight. Fig.~\ref{fig1}b presents ELA-measured trapped and precipitating electron energy fluxes in the outer radiation belt. The ratios of precipitating to trapped energy fluxes reach one at the sharp outer edge of the radiation belt in the bottom two panels, indicating broadband electron precipitation in the energy range of 50 keV up to 1 MeV. It is possible that the precipitation is a net result of many different scattering and acceleration mechanisms in the inner magnetosphere \cite{Sergeev93:scattering,Millan&Thorne07,Li&Hudson19}. Of particular interest is the complete absence of precipitation in the high-latitude ($\leq\sim$-70$^\circ$) region beyond the radiation belt. In the second panel, the plasma sheet precipitation with energies less than 300 keV seems to be confined to just outside the radiation belt within a narrow latitudinal extent of less than half a degree. This is potentially due to stretching of magnetic field lines and current sheet thinning during the substorm growth phase and prior to the injection \cite{Baker96,Runov21:thinning}.

Fig.~\ref{fig1}c shows that ELB traversed the magnetically conjugate region to MMS in the northern hemisphere during the injection period of 06:56--07:02 UT, although behind the injection front. In addition to intense and $>$500 keV electron precipitation in the outer radiation belt, ELB observed significant broadband electron precipitation below 500 keV from the plasma sheet region. Because the ELFIN spacecraft travels at a speed of $\sim$8 km/s and traverses a wide range ($\sim$15$^\circ$) of magnetic latitudes in the auroral ionosphere within $\sim$4 min, the precipitation observed by ELB represents mostly spatial features. As the injection and waves had been observed since 06:30 UT, we expect the energetic precipitation also took place before 06:56 UT, when ELB was not at the right position to capture it. The plasma sheet precipitation is identified according to (i) the precipitation region is observed poleward of the outer radiation belt boundary (or the isotropic boundary) with magnetic latitudes$>\sim$70$^\circ$ and $L_{IGRF}$ $\geq\sim$8 Re; (ii) the energies of plasma sheet electron precipitation are mostly $<$500 keV; and (iii) a clear energy flux decrease can be identified outside the radiation belt (with energies $>$500 keV). 

The identified intense plasma sheet precipitation during the injection by ELB is in sharp contrast to little precipitation observed by ELA before the injection occurred (Fig.~\ref{fig1}b). The upper energy limit of the high-latitude precipitation is roughly consistent with that of injected electrons as observed from MMS-1 (Fig.~\ref{fig1}a). Such precipitation may be explained by wave-particle interactions, or by field-line curvature (FLC) or current sheet scattering from the equatorial magnetosphere \cite{Sergeev93:scattering,Sergeev18:grl}. FLC scattering in principle produces relatively isotropic precipitation \cite{Sergeev93:scattering} and is more efficient for higher-energy electron precipitation than lower-energy electrons because of smaller ratios of curvature radius over electron gyroradius for higher-energy electrons. This ratio determines the efficiency of FLC scattering \cite{Buchner&Zelenyi89}. Smaller curvature-radius-to-gyroradius ratios will produce larger precipitating-to-trapped flux ratios for higher-energy electrons.

\begin{figure}
\centering
\vspace*{-0.5cm}\hspace*{-1.0cm}\includegraphics[width=1.1\textwidth]{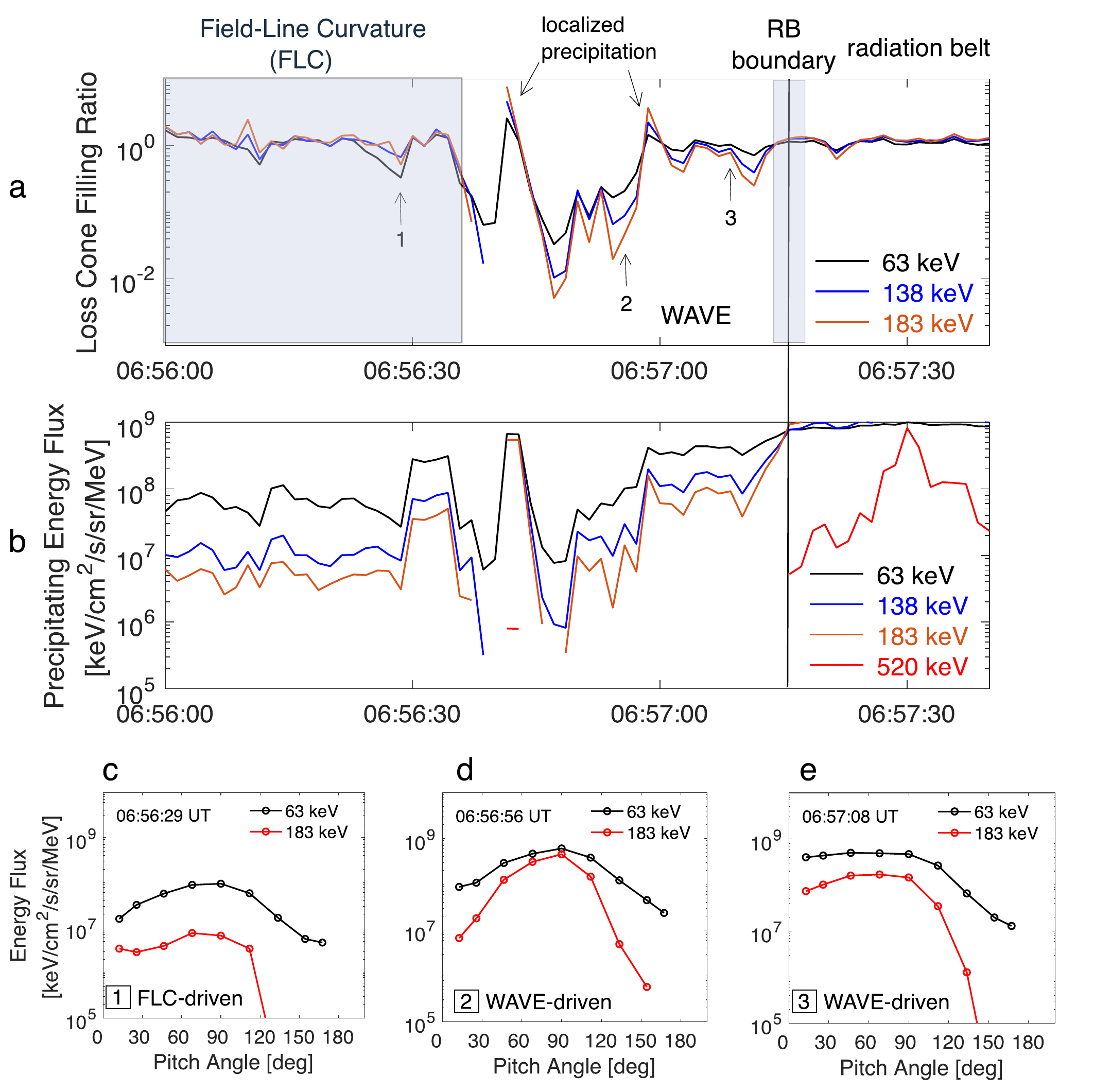}
\caption{(a) ELB-measured loss cone filling ratios for 63 keV, 138 keV, and 183 keV. Plasma sheet precipitation originated from field-line curvature (FLC) scattering and potentially from waved-driven scattering is indicated outside the radiation belt (RB). The isotropic ratios within the radiation belt are affected by flux saturation during this event. Two intervals show ratios larger than 1 due to the presence of localized (less than $\sim$10 km) sub-spin ($<$1.5 s) precipitation bursts. (b) Precipitating electron fluxes for 63 keV, 138 keV, 183 keV, and 520 keV. A sharp flux decrease is observed at the outer radiation belt boundary. There is little 520-keV electron flux beyond the identified radiation belt. (c) Three example electron pitch-angle distributions measured at 63 keV and 183 keV energy channels}. 
\vspace*{-1.3cm}
\label{fig2}
\end{figure}

Fig.~\ref{fig2} presents a closer view of the loss cone filling ratios and precipitating fluxes measured at 63 keV, 138 keV, and 183 keV, along with the identified wave-driven precipitation regions from ELB by comparing the loss cone filling ratios at three different precipitating energies. As shown by the example electron pitch-angle distributions at 63 keV and 183 keV, FLC-scattering is mostly associated with larger precipitation-to-trapped flux ratios at higher energies (Fig.~\ref{fig2}c), whereas potential wave-driven scattering is associated with larger precipitation-to-trapped flux ratios at lower energies (Fig.~\ref{fig2}d and~\ref{fig2}e). At relatively higher-latitude region during 06:56:00--06:56:35 UT, the precipitation is likely mapped to distant magnetotail regions where the magnetic field strength is weak, and the field-line curvature radius is small enough such that it is comparable to the energetic electron gyroradius. When the ratio of curvature-radius-over-gyroradius becomes smaller than $\sim$8 \cite{Sergeev83,Buchner&Zelenyi89}, effective electron precipitation driven by field-line curvature scattering is more likely to take place. This precipitation was thus observed by ELFIN in the relatively higher-latitude region. During 06:56:30-06:57:15 UT, as the spacecraft flew into relatively lower-latitude ionospheric regions where the corresponding equatorial magnetic fields are stronger and more dipolarized (i.e., $B_z$ increases and $\partial B_{x}/z$ decreases), the magnetic field curvature radius is increased \cite{Sergeev83,Lukin21}. Thus, we do not see precipitation driven by field-line curvature scattering in this region. Two localized (less than $\sim$10 km) sub-spin ($<$1.5 s) precipitation bursts with the precipitating-over-trapped flux ratios significantly exceeding 1 were observed by ELB during 06:56:35--06:57:10 UT. The flux ratios for these sub-spin bursts cannot be used to infer realistic loss-cone filling ratios due to spatial aliasing. Whether these precipitations can be attributed to wave-particle interactions or very localized curvature scattering remains unclear.

From the perspective of equivalent ionospheric currents (EICs) and spherical elementary current amplitudes (SECAs) inferred from the ground, the transient tailward flows and broadband waves encountered by MMS near 06:32 UT were associated with a sudden enhancement of the substorm current wedge (SCW) \cite{McPherron73,Shiokawa97,Kepko14:ssr} in Fig.~\ref{fig1}d. The injection and dipolarization observed by MMS near 06:35 UT were correlated with a transient, equatorward moving small-scale SECAs and EICs that swept across the MMS and ELB footprints in the panel d2 in Fig.~\ref{fig1}d. This magnetospheric injection can be reasonably mapped to the ionosphere and was associated with ELB-observed precipitation region later near 06:56 UT shown as the thickened green line. We use the TS04 storm-time model \cite{Tsyganenko&Sitnov05} coupled with the IGRF model \cite{Alken21:IGRF} to perform field-line mapping of MMS and ELB. The model inputs are constrained by real-time observations of the $Dst$ index (on average $\sim$-31 nT), IMF $B_y$ (on average $\sim$1 nT) and $B_z$ (on average $\sim$0 nT), and the solar wind density (on average $\sim$2 cm$^{-3}$) and speed (on average $\sim$650 km/s). The uncertainties associated with field-line mapping of MMS to the ionosphere ($\sim$100 km altitude) are estimated to be $\sim$2$^\circ$ in geographic latitude and $\sim$3$^\circ$ in geographic longitude by using different Tsyganenko models. Although the magnetic footprint MMS is only approximate during the substorm, it can be reasonably associated with the downward current region during the injection. Because the magnetic field configuration did not change appreciably after the initial injection in Fig.~\ref{fig1}a and because large-scale currents near the MMS footprint are very similar in location during the ELB crossing, dynamic field line mapping in panel d2 near 06:34:40 UT is similar to mapping during the ELB passage in panel d3. Although there exists a near 20-min time separation between the injection front and the ELB crossing, which is comparable to the injection passage time, the injection is persistent and develops into pile-up or overshooting of injection/dipolarization fronts at the near-Earth region, such that the fronts rebound and oscillate at the equator \cite{Panov10,Birn11,Schmid11:DF}. This dynamic oscillatory behavior has been observed in the SECA and EIC current system in our case. The dynamic movie of the current maps has been provided in Supporting Information.

\begin{figure}
\centering
\vspace*{-0.5cm}\hspace*{-1.0cm}\includegraphics[width=1.0\textwidth]{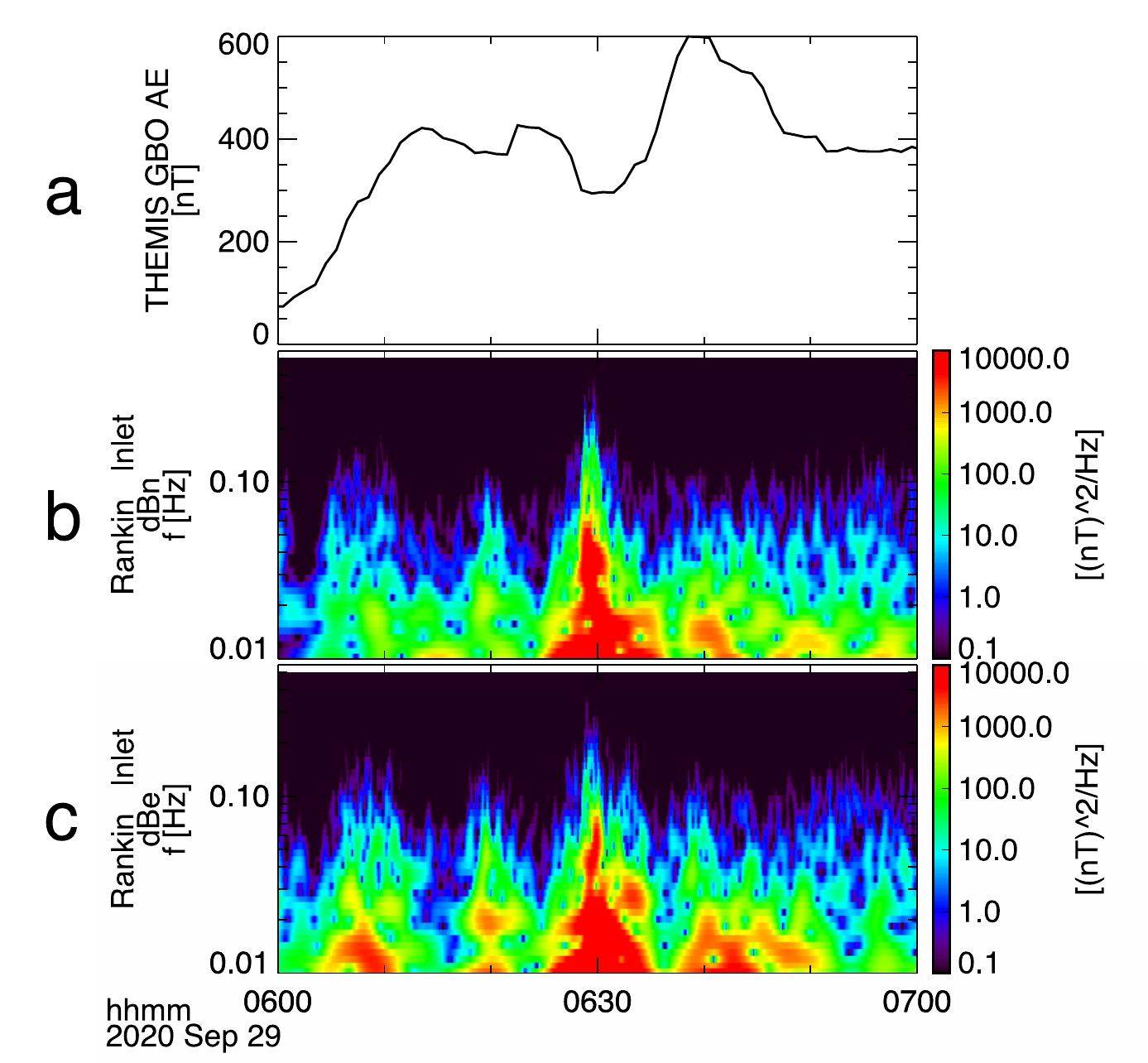}
\caption{(a) Magnetic $AE$ index measured from THEMIS ground-based observatories. (b, c) magnetic spectrograms obtained from the magnetometer stations at Rankin Inlet (RANK, at geographic latitude 62.82$^\circ$ and longitude 267.89$^\circ$). Results are shown for RANK in the parallel (or northward, dBn) and perpendicular (or eastward, dBe) components.}. 
\vspace*{-1.3cm}
\label{fig3}
\end{figure}

In addition to the current map, ground-based magnetometers conjugate to the ELFIN-precipitation region provide additional support that ELB-observed precipitation was likely associated with KAWs and the magnetospheric injection/dipolarization. Fig.~\ref{fig3} displayed the AE indices during the event along with dynamic magnetic spectra measured from the fluxgate magnetometer at Rankine Inlet and from the induction coil magnetometer at Fort Churchill stations. The locations of these two stations are displayed in Fig.~\ref{fig1}d relative to the ELB orbital footprints and precipitation. Fig.~\ref{fig3} demonstrates that enhanced broadband compressional waves below $\sim$1 Hz (shown in the northward component), known as Pi1B magnetic pulsations, were correlated with both the substorm onset near 06:30 UT and with the injection/dipolarization observed by MMS near 06:40 UT. Previous studies have established that compressional ULF waves in the Pi1-2 range are inherent features associated with substorm dipolarizations and ion fast flows \cite{Shiokawa97,Kepko14:ssr}. \citeA{Lessard06,Lessard11} have reported a similar correspondence between broadband compressional waves, substorm onset/dipolarizations, low-altitude dispersive Alfv{\'e}n waves, and Alfv{\'e}nic electron acceleration. Our observations of conjugate compressional waves suggest that ELFIN precipitation was potentially associated with kinetic Alfv{\'e}n wave activities in the magnetospheric plasma boundaries, such as injections fronts and plasma sheet boundary layers. In these regions, KAWs can be generated through mode conversion \cite{Hasegawa&Chen75,Lin12} or phase mixing \cite{Allan&Wright00}. A high correlation of KAWs with injections, dipolarizations, and fast ion flows has also been reported by many previous studies (see, e.g., \citeA{Chaston12:fast_flow}, \citeA{Ergun15}, and \citeA{Malaspina15}). 

It is worth emphasizing here that the conjugacy between MMS and ELB observations is established through the following procedures: (i) we first associate the MMS-observed injection and dipolarization with those ground-based observations of the currents of EICs and SECAs; and (ii) assume the MMS-observed injection fluxes and waves have similar characteristics to those in equatorial source regions; (iii) then, we use the TS04 model to perform field line mapping to locate the approximate footprints ($\sim$100 km) of MMS relative to the large-scale current systems during the injection and dipolarization. This relative position is valid within uncertainties of field-line mapping; (iv) ELB-observed precipitation region can be more reliably mapped to the ionosphere ($\sim$100 km) relative to the injection currents because ELB was at an altitude of $\sim$500 km and the magnetic field configuration close to the Earth can be modeled with little uncertainty by the IGRF model; (v) the ground-based magnetometer observations of compressional and shear Alfvén waves support that ELB-observed precipitation observations are probably linked to kinetic Alfv{\'e}n waves in the magnetosphere where MMS were located nearby. Therefore, the linkage of ELB precipitation to MMS is established through the corresponding EICs and SECAs of the injection and the ground-based wave observations. 

Because the MMS spacecraft were off the equator, we infer the equatorial magnitude of the dipolarization front associated with the injection based on ground-based $\Delta H$ measured by the mid-latitude stations, which are relatively unaffected by ionospheric currents and are known signatures of the field dipolarization during substorms \cite{Kokubun&McPherron81,Huang04}. Three ground-based magnetic perturbation measurements at middle latitudes are provided in Supporting Information. The inferred magnitude of dipolarization has a very typical value of $\Delta B_z$ $\sim$25 nT \cite{Runov11pss} and will be used to specify magnetospheric $B_z$ gradients in the following test particle simulations.

\begin{figure*}
\centering
\vspace*{-0.5cm}\hspace*{-1.0cm}\includegraphics[width=1.1\textwidth]{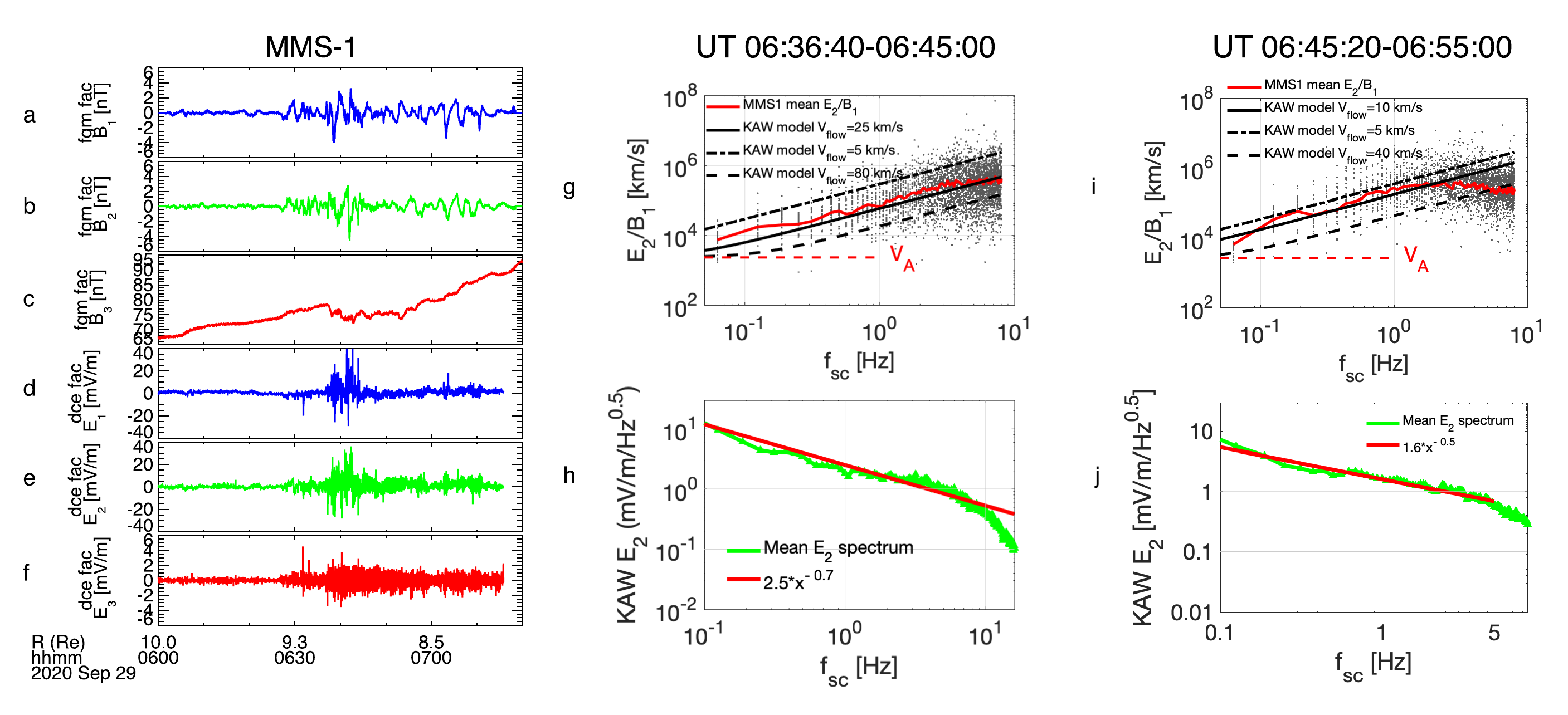}
\caption{(a-f)MMS-measured magnetic field and electric field in the field-aligned coordinates ($B_3$ is in the ${\bf B}$ direction), spanning the period from 06:00 UT to 07:20 UT. (g,i) Mean $E_2$/$B_1$ spectra (red) in comparison with the prediction by kinetic Alfv{\'e}n wave dispersion relation using different observed values of ion flows (black lines). The comparison is performed with data during 06:36:40--06:45:00 UT and 06:45:20--06:55:00 UT. The black dots represent the measured $E_2$/$B_1$ spectra, each calculated using a window size of 16 s. The local Alfv{\'e}n speed is shown as the red dashed line. (h,j) Least-square power-law fitting of the mean KAW $E_2$ spectra for the two time periods.}
\vspace*{-1.cm}
\label{fig4}
\end{figure*}

Fig.~\ref{fig4} presents the nature of kinetic Alfv{\'e}n waves associated with the injection observed by MMS-1 during the period of 06:00--07:20 UT. The electric field measurements demonstrate small-scale fluctuations and intermittent spikes. These small-scale fluctuations are quasi-electrostatic and relatively less evident in the magnetic field data. These features are consistent with the quasi-electrostatic property of KAWs with $k_\perp \gg k_{||}$ \cite{Hasegawa&Chen75}. We have transformed the measured fields into the field-aligned coordinates as mentioned above. We compare the measured $E_2/B_1$ spectra with the theoretical prediction of KAW dispersion relation, assuming the measured spacecraft frame spectra are largely due to Doppler shifts of KAW perpendicular wave structures due to ion flows \cite{Stasiewicz00,Chaston15:kaw}:
\begin{eqnarray}
\left|\frac{E_2}{B_1}\right|=v_A (1+k_{\perp}^2 \rho_{i}^2)\left[1+k_{\perp}^2(\rho_{i}^2+\rho_{s}^2)\right]^{-1/2},
\label{scaling}
\end{eqnarray}
where based on MMS observations of a background magnetic field $B_0$ $\sim$75 nT, an average (proton-dominated) ion number density $n_i$ $\sim$0.5 cm$^{-3}$, $T_{e} \simeq T_{i}$ $\sim$4 keV, the local Alfv{\'e}n speed $v_A=B_0/\sqrt{\mu m_i n_i}$ $\sim$2,300 km/s, $k_{\perp}\simeq2\pi f_{sc}/v_i$ is the KAW perpendicular wavenumber inferred from the spacecraft frame frequency $f_{sc}$ and perpendicular ion flows $v_i$ (up to 80 km/s and on average $\sim$25 km/s near the injection front based on MMS FPI observations), and $\rho_i$,$\rho_s$ are the corresponding ion thermal gyroradius and ion acoustic gyroradius. Following \citeA{Chaston12:fast_flow} and \citeA{Malaspina15}, we can test the assumption of Doppler-shift effects by examining the invariance of magnetic field spectra with $f_{sc}/|v_{i}|$. Because $f_{sc}/|v_{i}|=f/|v_{i}|+\vec{k} \vec{v_{i}}/2\pi|v_{i}|$, only the Doppler shift term is invariant with $|v_{i}|$. If the magnetic field spectrum as a function of $f_{sc}/|v_{i}|$ is also invariant with $|v_{i}|$, then the assumption of $f_{sc}\simeq k v_{i}/2\pi$ is reasonable. Such testing is provided in Supporting Information and shows that the Doppler-shift assumption is mostly justified with $v_i$ larger than $\sim$25 km/s.

The polarization predicted from KAW dispersion in Eq~\ref{scaling} assumes plane waves in a uniform medium with purely H$^+$ ions. The full dispersion relation from \citeA{Lysak&Lotko96} and \citeA{Lysak08} is expanded in the limit of  $m_{e}/m_{i}\ll \beta_{e}<1$ and $f\ll f_{ci}$ for Eq~\ref{scaling}, where $\beta_e$ and $f_{ci}$ are the electron plasma beta and ion cyclotron frequency. Application of Eq~\ref{scaling} to plasma sheet observations of kinetic Alfv{\'e}n waves has been examined first by \citeA{Wygant02} using Polar observations and on a larger database by \citeA{Chaston12:fast_flow} using THEMIS observations. 

Near the front before 06:45 UT, the measured wave fields are consistent with those predicted by the KAW model, when applying an average transverse ion flow velocity of 25 km/s measured by MMS (Fig.~\ref{fig2}g). The variations of $E_{2}/B_{1}$ spectra may be attributed to the variations of ion flow velocities during this interval, or due to nonlinear effects associated with small-scale Alfv{\'e}n waves (e.g., \citeA{Wygant02}). The calculated $E_{2}/B_{1}$ spectra (fast mode data) have frequencies up to 8 Hz, corresponding to $k_\perp \rho_{i}$ $\sim$3-270 in the kinetic branch \cite{Lysak&Lotko96}. In addition to dispersion fitting, we have also performed coherence analyses of $E_2$ and $B_1$ measurements. The data are consistent with travelling Alfv{\'e}n waves at frequencies below $\sim$5 Hz. This information has been provided in Supporting Information.

Fig.~\ref{fig2}h presents the fitted power-law spectrum of the mean $E_{2}$ field as $E_{f}=E_0 f_{sc}^{-\nu}$ (mV/m/$\sqrt{Hz}$), where $E_0=$2.5 mV/m and $\nu=$0.7, an observable to be used in the test particle simulations. Similarly during the period of 06:45:20--06:55:00 UT, only 1 min before ELB traversed the MMS footprint in Fig.~\ref{fig1}d, the KAW spectra show consistency with the KAW dispersion relation in frequencies up to $\sim$2 Hz. The fitted power-law spectrum of the mean $E_{2}$ field has $E_0=$1.6 mV/m and $\nu=$0.5. Because MMS observations were not at the equator within the center of the fast ion flow channel, larger KAW amplitudes may be associated with wave particle interaction processes responsible for ELB-observed precipitation. We will mainly use the observed injection-front KAW intensities but also apply different wave amplitudes for the test particle simulations to explore potential variations in scattering rate.

\section{Resonant interaction between KAW and injected electrons}

We consider a scenario in which injected electrons interact with kinetic Alfv{\'e}n waves (KAWs) in the magnetic field $B_z$ gradients along the $x$ direction (e.g., in GSM coordinates) near equator. Including energetic electron perpendicular magnetic drifts (i.e., $\nabla B$ drifts) associated with $B_z$ field gradients, the wave-electron resonant condition is given as \cite{Summers98}:
\begin{eqnarray}
\omega - k_\perp v_{drift} - k_{||} v_{||} = n \Omega_{ce}/\gamma,
\label{resonance}
\end{eqnarray}
which is simplified as $v_{||}\simeq -k_\perp v_{drift}/k_{||}$, where the KAW frequency $\omega$ is negligibly small, $n=0$, corresponding to Landau resonance. As shown in Supporting Information, KAW real frequency $\omega\ll k_\perp v_i$ where $v_i$ is on the order of 50 km/s. With an equatorial magnetic field diplarization of 25 nT and a typical gradient scale of 400 km on the order of the local ion gyroradius in the magnetotail \cite{Runov11pss}, $v_{drfit}$ is generally larger than 100 km/s for 50 keV electrons at above 10$^\circ$ pitch angle. Therefore, the KAW real frequency will typically be negligibly smaller than the Doppler-shift term. However, when electron pitch angles decrease to near the loss cone, the real frequency term cannot be neglected. We consider resonant electrons move in the direction opposite to the KAW parallel wave vector $k_{||}$ direction whereas the $\nabla B$ drift of resonant electrons is aligned with the perpendicular wave vector $k_\perp$ direction. The resonant energy and efficiency of electron scattering are collectively determined by KAW intensities, wave normal angles, and electron perpendicular magnetic drifts.   

\section{Test particle simulations and precipitating flux comparison}

We use a test particle simulation code to estimate electron pitch angle scattering by broadband electric fields of kinetic Alfv{\'e}n waves associated with injections and magnetic field $B_z$ gradients. We solve full relativistic Lorentz equations of electrons and obtain the electron position ($\vec{r}$) and momentum ($\vec{p}$) using the classic 4th order Runge-Kutta integrator (see tests in \cite{Shen&Knudsen20}). The relativistic electron equation of motion is: 
\begin{eqnarray}
  \frac{d\vec{p}}{dt}=q_e \left[\vec{E}+\frac{\vec{p}}{m_e \gamma} \times \vec{B}\right]
  \label{resonance}
\end{eqnarray}
where $\vec{p}=\gamma m_e\vec{v}$ is the electron momentum, $\gamma=[1+p^2/(m_e c)^2]^{1/2}$, and the magnetic field is specified as $\vec{B}=B_z(x)\;\hat{z}=B_0 [1.1+0.9\tanh{(x/L_x)}]/2\;\hat{z}$, where $B_0$ $\sim$25 nT and $L_x$ $\sim$ 400 km based on current and previous observations as aforementioned \cite{Runov11pss}. 

Note that we have assumed no magnetic field variations in the $z$ direction and have neglected the full bounce motion along the field line and focused on local equatorial wave-particle interactions. Thus, the interaction near the equator is artificially prolonged and exaggerated and the interaction along the field line is weakened due to lack of bounce motion. However, because interactions between KAWs and electrons mostly occur near the equator where perpendicular magnetic drifts are the most significant, the limitation of neglecting bounce motion does not negate our key results shown in the following, i.e., energetic injection electrons below $\sim$500 keV can be driven into the loss cone by KAWs through Doppler-shifted Landau resonance. We also only consider the electric field spectra of KAW and neglect the small magnetic perturbations in the parallel direction \cite{Hollweg99}. These magnetic perturbation effects and electron bounce motion will be examined in a future study.

The electric fields of KAWs are specified as $\vec{E}=E_\perp \hat{y}- E_{||} \hat{z}$, in which we have \cite{Stasiewicz00}:
\begin{eqnarray}
    E_\perp=\sum_{k_\perp \rho_i=2-128}^{}E_k\;\cos{\left(k_{||}z-k_\perp y-\omega t + \phi_{rand}\right)},
\end{eqnarray}
\begin{eqnarray}
    \frac{E_{||}}{E_\perp}=-\frac{k_{||} k_\perp \rho_s^2}{1+k_\perp^2 \rho_i^2}\simeq-\frac{k_{||}}{4k_\perp}.
\end{eqnarray}
where $\omega$ is set to a small number $2\pi\times 0.05$, $E_\perp$ and $E_{||}$ are the perpendicular and parallel components of the KAW broadband ($k_\perp \rho_i=$2--128) electric fields. In order to single out wave effects, no background DC electric fields are included. The perpendicular magnetic field perturbations are implicitly included through the parallel electric fields based on Faraday's law.

In simulations, the broadband KAW electric fields have 2,450 wavenumbers with (i) a representative spectrum $E_k=E_0\left(k_\perp v_i/2\pi\right)^{-\nu}$, where $\nu$ is $-0.7$ based on the measured spectrum in Fig.~\ref{fig4}h, and $E_0$ is $0.4$ mV/m as reduced to conserve total wave power for increased $k$-modes in the frequency range of 0.1--6.3 Hz, and (ii) a stepsize of 0.05 for $k_\perp \rho_i$ $\sim$2--128 to ensure stochastic interaction between electrons and KAWs and the absence of artificial nonlinear Landau resonance trapping for individual $k$-modes introduced by sampling of the spectrum  \cite{Karney78,Karimabadi90:waves}. When converting the $k$ spectrum to the frequency spectrum, we have specified a typical ion flow velocity $v_i\;$=50 km/s, two times the average value used for fitting in Fig.~\ref{fig4}g. This allows including a larger portion of the measured spectrum while keeping the number of $k$ numerically manageable and the stepsize small enough to avoid artificial nonlinear effects due to sampling, albeit at the expense of reducing the measured electric field amplitudes by a factor of 1$-$2$^{-0.7}=$0.4. The variation in amplitude will be accounted for by testing different KAW $E_0$.

\begin{figure}
\centering
\vspace*{0.1cm}\hspace*{-1.0cm}\includegraphics[width=1.0\textwidth]{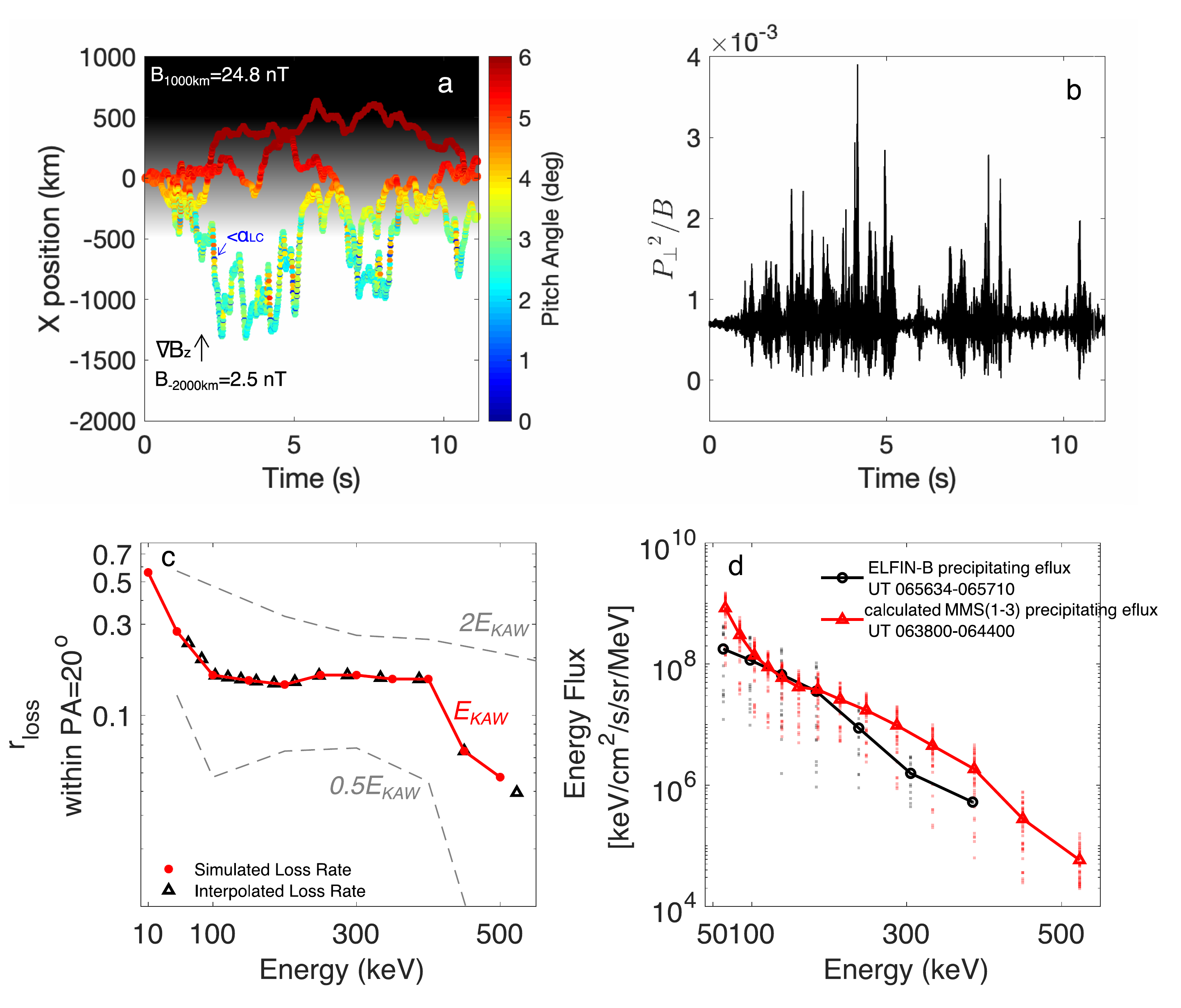}
\caption{(a) Three test electron trajectories and pitch angle variations as a function of time and position in $x$ direction, in which the magnetic field $B_z$ gradients are present. (b) Normalized magnetic moment variations of the three test electrons with an arbitrary unit. (c) Test particle simulation of electron loss rates ($r_{loss}$, red stars) driven by kinetic Alfv{\'e}n waves and magnetic field gradients. Examples of initial and lost electron pitch-angle and energy distributions and the loss rate calculation for several different energies are provided in Supporting Information. The black triangles are interpolated loss rates at energy channels measured by MMS, in order to calculate precipitating electrons within the loss cone from observations outside the loss cone by MMS. The two grey-dashed lines indicate the obtained loss rates when we double and halve the KAWs electric fields. (d) Comparison of precipitating electron energy fluxes measured by the three MMS spacecraft (1--3) in the magnetosphere and by ELFIN-B in the ionosphere when only wave-driven precipitation intervals are used. }
\vspace*{-0.5cm}
\label{fig5}
\end{figure}

Fig.~\ref{fig5}a and~\ref{fig5}b present the evolution of the trajectories (in $x$ direction) and magnetic moments of three test particles with an initial pitch angle of 5$^\circ$ and an energy of 100 keV. The three electrons are initiated from the center ($x=0$) of the gradient magnetic field with both random gyrophases and random locations in the $y$ position ($|y_0|\leq\rho_i$). The electron orbits are integrated for a period of $\sim$11 s with a stepsize $dt=$1/400$f_{ce}$. Fig.~\ref{fig5}a shows that as electron drifts towards a weaker $B_z$ field region, its pitch angle decreases; as electron drift towards a stronger field region, the pitch angle increases. These variations of pitch angle are due to adiabatic transport across the $B$ field gradients on the large time scale, as shown by the steady baseline of Fig.~\ref{fig5}b. This transport is a result of small but accumulative $E\times B$ pushes in the $x$ direction. In this case, no $E$-field induced acceleration/deceleration occurs in the electron guiding-center frame. 

Fig.~\ref{fig5}b shows that intermittent pitch angle and momentum alterations also take place, especially when electrons move towards the edge of the weaker field region. These sudden and intense variations of pitch angle and momentum lead to sporadic loss of electrons ($\alpha\leq$2$^\circ$), which can be attributed to two processes: (i) KAW $E_y$ fields contain spatially small-scale, intense fluctuations, so that acute changes of $E\times B$ drifts take place, which transforms into sudden electron acceleration or deceleration; (ii) electron $\nabla B$ drift velocities vary significantly as electrons approach the edge of the weaker field region, so their perpendicular momenta change appreciably. The above two factors work together and produce electron pitch angle variations on a much shorter time scale than those driven by adiabatic transport. This is evidently shown as the momentum spikes in Fig.~\ref{fig5}b. As a result, electrons can be driven into the loss cone by the combination of the large-scale adiabatic transport and small-scale momentum kicks due to varying $E\times B$ and $\nabla B$ drifts. 

At the injection front, MMS observed counter-streaming electrons with a plateau pitch angle distribution within $\pm$45$^\circ$. The average electron pitch-angle distribution measured by MMS is provided in Supporting Information. Through test runs for electrons in the energy range of 10--500 keV, we find that only electrons with pitch angles less than 20$^\circ$ can be driven into the loss cone for the given KAW spectrum and gradient magnetic field. To calculate the loss rate applicable to observations, we specify a uniform pitch angle distributions below 20$^\circ$, comprising discrete pitch angle elements ($\delta$ functions) of 5$^\circ$, 10$^\circ$, 15$^\circ$, and 20$^\circ$, each with 100 test electrons. Electrons are counted as being lost if their pitch angles decrease to be smaller than 2$^\circ$ during the integration period of $\sim$11 s. The loss rate is obtained as $r_{loss}=N_{loss}/400$. This procedure is repeated for different electron energies. Test particle simulation examples of electron pitch-angle and energy variations as well as loss rate calculations for different energies are provided in Supporting Information.     

Fig.~\ref{fig5}c presents the calculated electron loss rates for the observed energy range of the injection and precipitation. The loss rates reach more than 50\% for 10 keV electrons but drop to near 0\% for $>$500 keV electrons. The extent of pitch angle variations and the chance of loss largely depend on the amount of induced perpendicular momentum variations, which are limited by the model $E$ fields and magnetic field gradients. Therefore, for a given scale of momentum variations, lower-energy electrons will experience proportionally larger pitch angle decreases. To account for the uncertainty in KAW amplitudes associated with precipitation in our observations, Fig.~\ref{fig5}c also presents variations in loss rate when we multiply the observed amplitudes of KAWs electric fields by a factor of 2 (as 2$E_0$) and 0.5 (as 0.5$E_0$). When the electric field amplitudes become weaker, scattering may become relatively more effective for higher-energy electrons in a certain energy range. This is because the resonant energy at the same pitch angle increases with smaller $k_\perp$ (or smaller wave normal angles \cite{Chaston09}) for a given field model, if we recognize the energy dependence of $v_{drift}$ in Eq.~\ref{resonance}. This may produce flattened loss rates, because when the effects of electric field amplitudes become weaker, the falling shape of the wave spectrum determines the relative scattering efficiency. Furthermore, although electron bounce motion has been ignored in our simulations, we expect that inclusion of bounce motion will increase the parallel electric field effects associated with KAWs, therefore the extent of pitch angle decreases near the equator will likely be strengthened, albeit with reduced overall scattering efficiency due to shorter dwelling time near the equator. Also, electron curvature drift effects have not been incorporated due to absence of magnetic field gradient in $z$ direction in our idealized field model. Inclusion of this curvature drift will likely enhance the total perpendicular magnetic drift, thus shifting the resonance to a smaller $k$ or larger electric fields regime for the same resonant energy. 


We apply the loss rates in Fig.~\ref{fig5}c to estimating precipitating electron energy fluxes from MMS in the magnetosphere in Fig.~\ref{fig5}d. We obtained the average electron parallel energy fluxes within 20$^\circ$ pitch angles from three MMS spacecraft (excluding MMS-4) when they were within the injection front, where magnetic field gradients were prominent (06:38--06:44 UT). The precipitating fluxes are calculated as the product of the observed parallel energy flux and the loss rate at the corresponding energy. Fig.~\ref{fig5}d compares the inferred precipitating energy fluxes from MMS with those measured by the ELFIN-B spacecraft, which captured wave-driven plasma sheet electron precipitation during the period of 06:56:34--06:57:10 UT (Fig.~\ref{fig2}). We have removed the contribution of backscattered electrons to the precipitating energy fluxes measured by ELFIN-B. In addition, the ELFIN-B flux data have been denoised using an uncertainty threshold of 50\% based on counting statistics. The resultant average precipitating energy fluxes measured by ELFIN-B (black line) and MMS (red line) are consistent at energies below $\sim$300 keV within spectral variations. The results in Fig.~\ref{fig5} suggest that the precipitating electrons below $\sim$300 keV observed by ELFIN-B in the plasma sheet region are probably driven by resonant wave-particle interaction between magnetospheric injection electrons and kinetic Alfv{\'e}n waves in the dipolarization front magnetic field gradients. 

\section{Discussion and Conclusions}

Previous studies have shown that kinetic Alfv{\'e}n waves (KAWs) are important for accelerating electrons mainly in the thermal energy range (several eV up to a few keV) when electrons have parallel velocities close to the local Alfv{\'e}n speed \cite{Wygant02,Watt&Rankin09:prl,Watt&Rankin12} and sometimes nonlinear resonance broadening occurs \cite{Artemyev15:jgr:KAW,Damiano15}. In more recent studies, standing waves of kinetic field line resonances (FLRs) have been suggested to drive pitch-angle scattering and radial diffusion of radiation belt electrons with energies above a few hundred keV via drift-bounce resonance \cite{Chaston18:dropout,Chaston18:scattering}. But the time scales of such scattering are $\sim$hours compared with an injection time period of up to tens of minutes.

In this study, we consider KAWs with $k_\perp \rho_i>$1 interacting with plasma sheet injection electrons via Doppler-shifted Landau resonances in the energy range of 10--500 keV, which is the main population of injected electrons from the magnetotail \cite{Gabrielse14, Turner16}. This range of resonant energies with KAWs has so far not been explored but can be significant if we include $\nabla B$ drifts in the magnetic field gradients associated with injections and dipolarizations, where kinetic Alfv{\'e}n waves have been found to be pervasive \cite{Chaston12:fast_flow,Chaston15:kaw,Huang12:angeo,Cheng21}. Using conjugate ELFIN and MMS spacecraft observations and interpreting these observations through test particle simulations, we have the following results:

\begin{enumerate}
\item We have reported direct observations of tens to hundreds of keV electron precipitation most likely driven by resonant interaction with KAWs in the magnetic field gradients associated with a magnetotail injection. 
\item The magnetic field gradients and the associated $\nabla B$ drifts allow Doppler-shifted Landau resonant interaction between injected electrons and KAWs, producing scattering and precipitation of injection electrons. Electron losses are attributed to a combining effect of large-scale adiabatic transport across the gradient magnetic field and small-scale perpendicular momentum kicks due to rapidly varying $E\times B$ and $\nabla B$ drifts.
\item Taking into account the estimated electron loss rates from simulations, the calculated precipitating electron energy fluxes from MMS are roughly consistent with plasma sheet energetic (50--$\sim$300 eV) precipitation observed by ELFIN-B.   
\end{enumerate}

\citeA{Clilverd08} and \citeA{Gabrielse19} have suggested that large-scale injections and dipolarizations during substorms are closely associated with energetic ($\geq$30 keV) electron precipitation as observed by ground-based riometers. Wave-particle interactions are thought to be necessary in that process. The mechanism of KAW-driven precipitation proposed in our paper may be a significant contributor to such electron precipitation.

\section{Open Research}
ELFIN data can be accessed through \url{ http://data.elfin.ucla.edu/}. MMS data can be obtained through \url{https://lasp.colorado.edu/mms/sdc/public/about/browse-wrapper/}. Matlab plotting code for simulation results is available through \url{https://doi.org/10.5281/zenodo.5728276}. We gratefully acknowledge the use of ground-based magnetometer data from SuperMAG (\url{https://supermag.jhuapl.edu/mag/}). EICs and SECAs datasets can be accessed from \url{https://doi.org/10.21978/P8D62B} and \url{https://doi.org/10.21978/P8PP8X}. Data analysis was done using SPEDAS V4.1, see \citeA{Angelopoulos19}.

\acknowledgments
Y.S., A.V.A., X.J.Z., and V.A. acknowledge support by NASA awards 80NSSC21K0729, NNX14AN68G, and NSF grants AGS-1242918, AGS-2019950, AGS-2021749, and 1914594. I.V. was supported by NASA Heliophysics Supporting Research grant No. 80NSSC20K1325. J.M.W. acknowledges NASA grants 80NSSC18K1227 and 80NSSC20K1364 and the NASA HPDE contract 80GSFC17C0018. We are grateful to NASA’s CubeSat Launch Initiative for ELFIN's successful launch in the desired orbits. We acknowledge early support of ELFIN project by the AFOSR, under its University Nanosat Program, UNP-8 project, contract FA9453-12-D-0285, and by the California Space Grant program. We acknowledge critical contributions of numerous volunteer ELFIN team student members.


%
%

%
%
%
%
%

\end{document}


%
%


\title{Supporting Information for ``Tens to hundreds of keV electron precipitation driven by kinetic Alfv{\'e}n waves during an electron injection"}
%
%

%
%



\authors{Yangyang Shen\affil{1}, Anton V Artemyev\affil{1,2}, Xiao-Jia Zhang\affil{1}, Vassilis Angelopoulos\affil{1}, Ivan Vasko\affil{2,3}, Drew Turner\affil{4}, Ethan Tsai\affil{1}, Colin Wilkins\affil{1}, James M Weygand\affil{1}, Christopher T Russell \affil{1}, Robert E Ergun\affil{5}, Barbara L Giles\affil{6}}

\affiliation{1}{Department of Earth, Planetary, and Space Sciences, University of California, Los Angeles, California, USA}
\affiliation{2}{Space Research Institute of Russian Academy of Sciences, Russia, Moscow}
\affiliation{3}{Space Science Laboratory, University of California, Berkeley, California, USA}
\affiliation{4}{Johns Hopkins University Applied Physics Laboratory, Maryland, USA}
\affiliation{5}{Laboratory for Atmospheric and Space Physics, University of Colorado Boulder, Boulder, CO, USA.}
\affiliation{6}{NASA Goddard Space Flight Center, Greenbelt, MD, USA.}


%
%

%

\begin{article}

%
%

\noindent\textbf{Supporting Information of Kinetic Alfv{\'e}n Wave Driven Electron Precipitation}
\begin{enumerate}
\item Movie M1
\item Figure S1
\item Figure S2
\item Figure S3
\item Figure S4
\item Figure S5
\end{enumerate}

\section*{Introduction} 

This Supporting Information provides a movie (M1) showing 2D dynamic maps of equivalent ionospheric currents (EICs) and spherical elementary current amplitudes (SECAs) inferred from a dense network of North American and Greenland ground-based magnetometers \cite{Amm&Viljanen99,Weygand11}. The current maps are calculated every 10 s with a spatial resolution of $\sim$350 km. Dynamic, oscillatory currents related to injections/dipolarizations can be observed during the period of 06:30--07:00 UT. 

We also include one figure (S1) to show the ground-based magnetic horizontal perturbations ($\Delta H$, or $dBn$ in the northward component) from three stations located at middle latitudes in our conjunction event, and one figure (S3) to show testing of the Doppler-shift assumption associated with kinetic Alfv{\'e}n waves by examining magnetic field spectra invariance with ion flows, one figure (S3) to show the coherence and phase relation between $E_2$ and $B_1$ signals which suggests travelling Alfv{\'e}n waves, one figure (S4) to show MMS-observed counter-streaming electrons with a plateau pitch angle distribution within $\pm\sim$45$^\circ$, one figure (S5) to show examples of initial and final lost electron pitch-angle and energy distributions for several different energies.

\section*{Ground-based Magnetic Perturbations} 
The ground-based magnetometer data are from the SuperMAG project, providing typically magnetic perturbation measurements at one sample per second time resolution \cite{Gjerloev12}. Data were obtained from the three middle-latitude stations of PIN ($lat\sim$50.20$^\circ$, $lon\sim$263.96$^\circ$), C04 ($lat\sim$50.06$^\circ$, $lon\sim$251.74$^\circ$), and T32 ($lat\sim$49.39$^\circ$, $lon\sim$277.68$^\circ$). The measured $\Delta H$ (or $dBn$ in the main text as the northward component) from these three stations are relatively unaffected by ionospheric currents and thus are used to infer the magnitude of dipolarization fronts associated with the injection \cite{Kokubun&McPherron81,Huang04}.

\section*{Testing for Doppler-shift Assumption of KAWs} 
Typically, frequencies of kinetic Alfv{\'e}n waves (KAWs) are much less than the proton cyclotron frequency (i.e., 1 Hz in our case). Therefore, these high-frequency (up to 8 Hz) field fluctuations may be assumed to be Doppler-shifted structures due to ion flows. This is possible because KAWs have $k_\perp\gg k_{||}$ so that the perpendicular wave phase speed can be sub-Alfv{\'e}nic \cite{Chaston12:fast_flow}. As discussed in the main text, we can test the assumption of Doppler-shift effects by examining the invariance of magnetic field spectra with $f_{sc}/|v_{i}|$, where $f_{sc}$ is the spacecraft frame frequency and $|v_i|$ is the magnitude of ion flows mainly in the perpendicular direction. If the magnetic field spectra as a function of $f_{sc}/|v_{i}|$ is also invariant with $|v_i|$, then the assumption of Doppler-shift effects is reasonable. Figure S2 shows that the magnetic field spectra variations as a function of ion perpendicular drift velocities (measured by the MMS1 FPI instrument during the corresponding intervals). The fitted magnetic field spectral indices converge to a constant of close to -0.8 as ion flow speed increases. The threshold velocity is identified to be $\sim$30 km/s when the width/variance of the scattered spectral indices decreases to be less than $\sim$0.5. The convergence indicates that $f_{sc}\simeq k v_i/2\pi$ is marginally justified by ion flow speeds of less than $\sim$30 km/s and justified for flow speeds above.  Similar tests and interpretations can be found in \citeA{Chaston12:fast_flow} and \citeA{Malaspina15}.

\section*{Test particle simulations of electron distributions and loss rates} 
As mentioned in the main text, electron motion is integrated for $\sim$11 s. Electrons are considered lost when their pitch angles decrease to below 2$^\circ$. We use the kinetic Alfv{\'e}n wave (KAW) amplitude spectrum with $E_0=$0.4 mV/m in the simulations to conserve wave power consistent with MMS observations while including 2,450 wavenumbers to avoid artificial nonlinear trapping due to sampling. The energies of lost electrons decrease due to perpendicular momentum decrease while interacting with spatially varying KAWs electric fields. Only the electrons with initial pitch angles less than 20$^\circ$ are pushed into the loss cone. The loss rate is therefore calculated as $r_{loss}=N_{loss}/$400, where $N_{loss}$ is the number of lost electrons from the population with initial pitch angles within 20$^\circ$.

\end{article}


%
%
%
%

\begin{figure}
\centering
\vspace*{0cm}\includegraphics[scale=0.6]{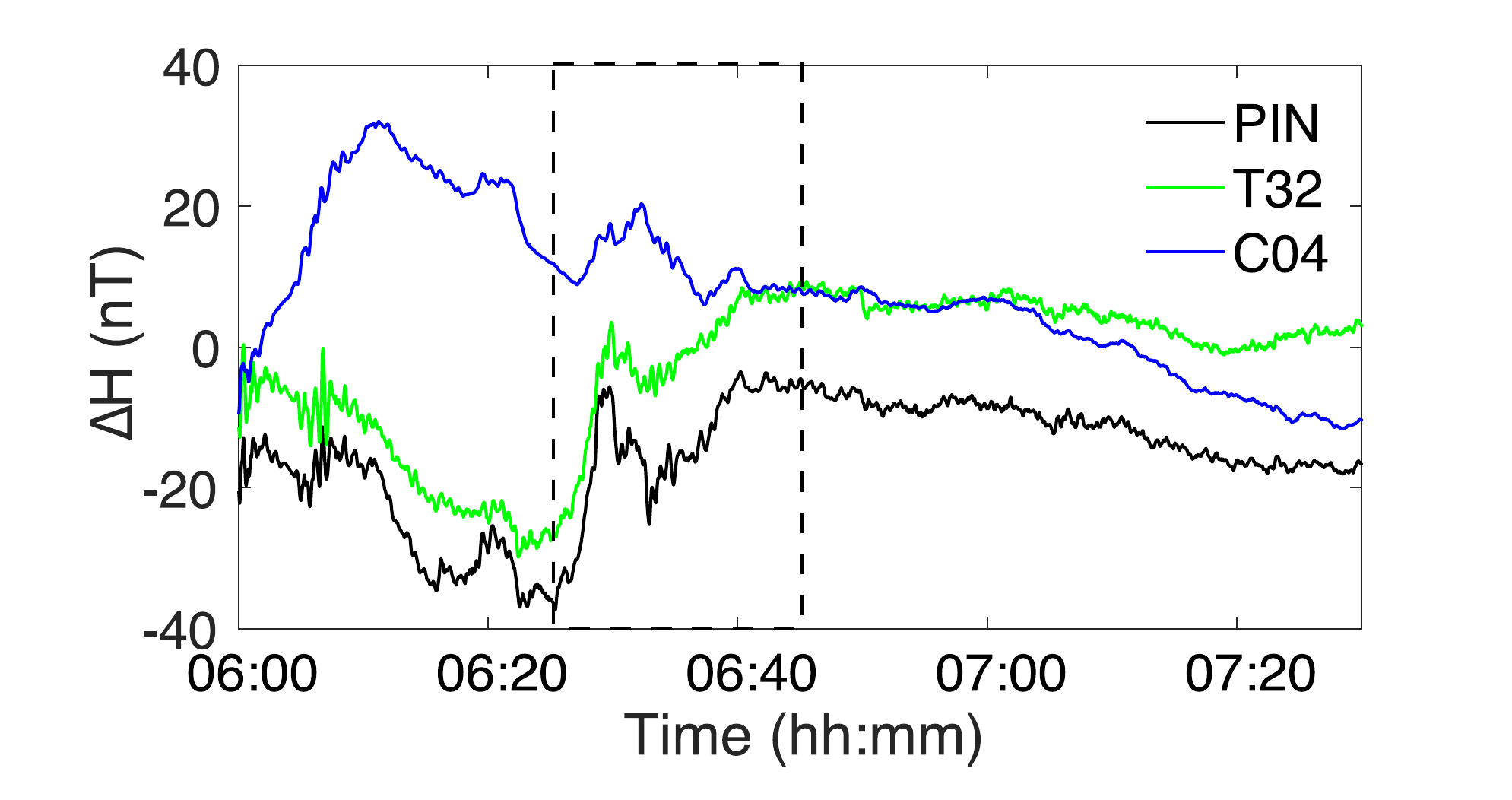}
\caption{Ground-based magnetic perturbations ($\Delta H$ or $dBn$, the northward component) observed from 3 magnetometer stations of PIN, C04, and T32 at middle latitudes. The time period in conjunction with MMS observations were highlighted by the dashed box region. The baselines of $\Delta H$ are provided as they are from background removal techniques described in a paper for SuperMAG \cite{Gjerloev12}. The measured magnetic perturbations are used to infer the magnitude of dipolarization fronts in the equatorial magnetosphere.}
\vspace*{0cm}
\label{figS1}
\end{figure}
\clearpage
\newpage

\begin{figure}
\centering
\vspace*{0cm}\includegraphics[scale=0.6]{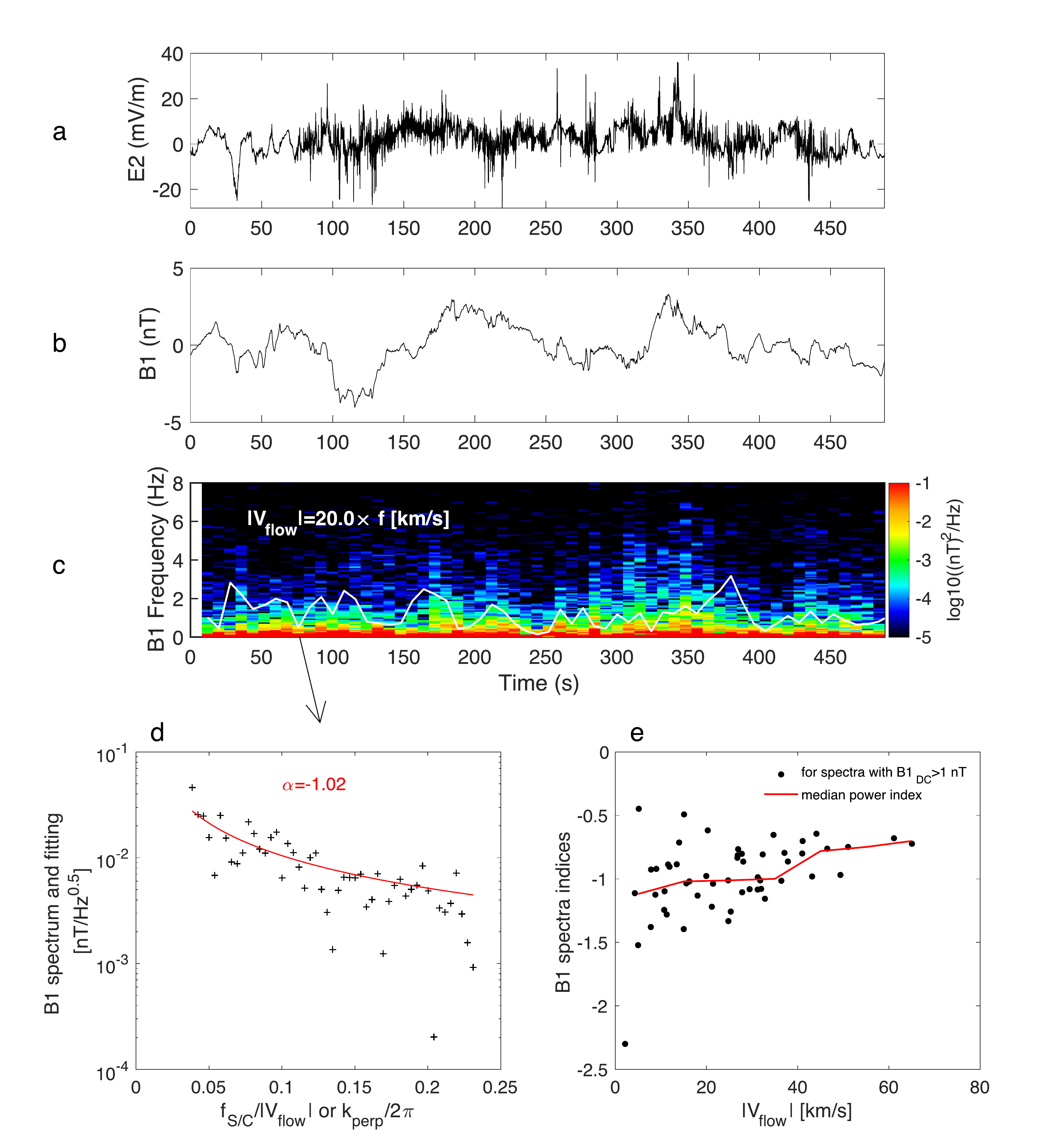}
\end{figure}
\begin{figure}
\caption{Testing the invariance of magnetic field spectra (as a function of $f_{SC}/|v_{i}|$) with different ion flow speeds ($|v_{i}|$ or $|V_{flow}|$) in the perpendicular-to-$B$ direction, during the interval of 06:36:40 to 06:45 UT (500 s). (a) Electric field $E_2$ variations (similar to $E_y$ in the GSM coordinates). (b) Magnetic field $B_1$ perturbations. (c) Magnetic field spectrogram, with Fourier transform applied during each time window of 16 s (weighted by the same size Hanning window and overlapped by 8 s for consecutive windows). The magnitudes of perpendicular ion flow velocities are calculated as $20\times f$ km/s and are overplotted as the white line. (d) $B_1$ spectral index ($\alpha$) least-square fitting during one of the intervals indicated by the black arrow. The spectrum varies as a function of $f_{SC}/|v_{i}|$. (e) Statistical distribution of the power-law indices derived for intervals when the DC $B_1$ component is larger than 1 nT to exclude noise background.}
\vspace*{0cm}
\label{figS2}
\end{figure}
\clearpage
\newpage

\begin{figure}
\centering
\vspace*{0cm}\includegraphics[scale=0.6]{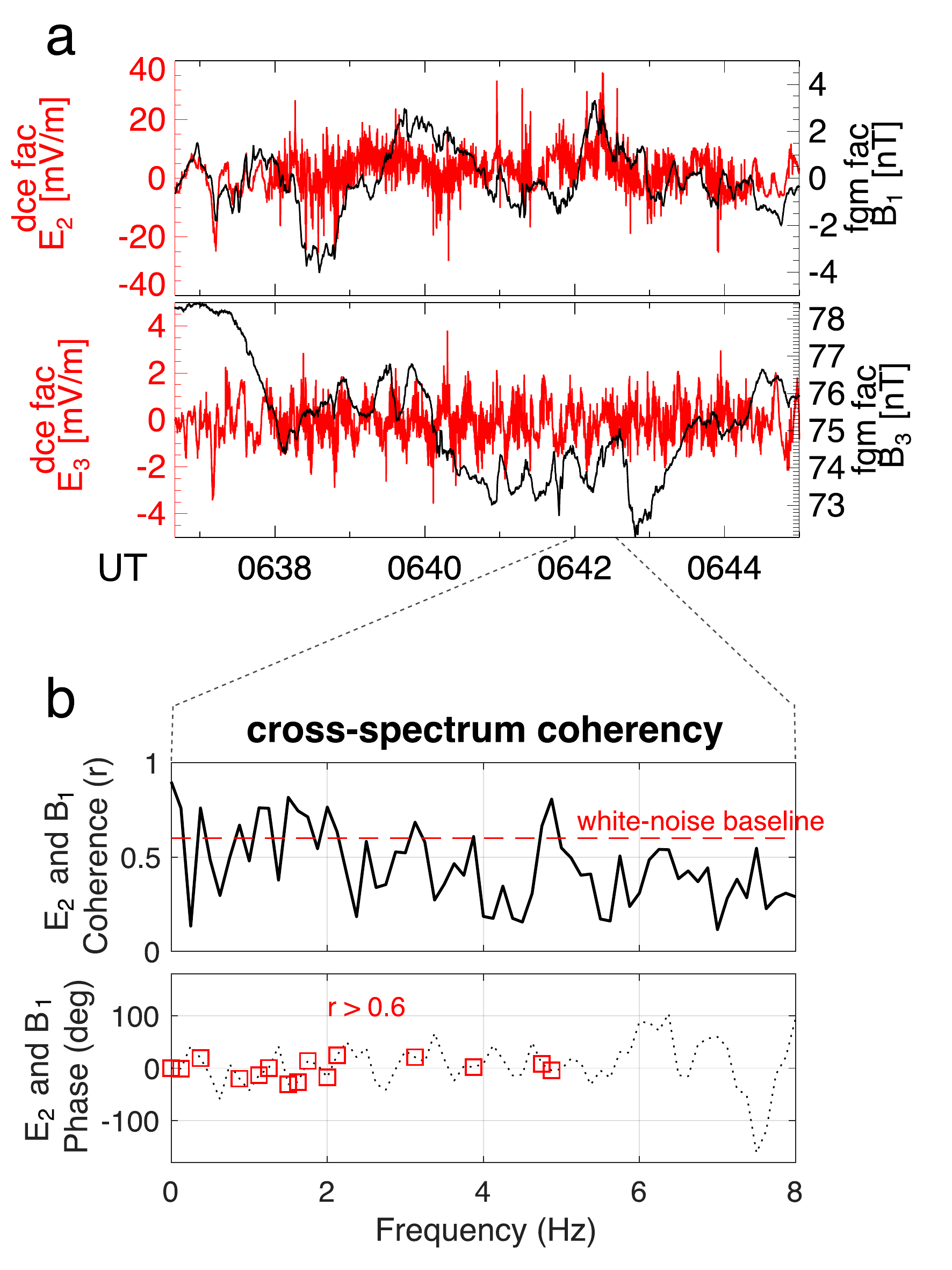}
\caption{(a) DC-coupled magnetic field $B_1$ (perpendicular) and $B_3$ ($B$-field-aligned) components in 16 samples per second (sps) and electric field $E_2$ (perpendicular) and $E_3$ (B-field-aligned) components in 32 samples per second time series data during the period of 06:36:40-06:45:00 UT. We see roughly correlated electric and magnetic field perturbations. (b) Cross-spectrum coherence ($r$) and phase relations between $E_2$ and $B_1$ examined during the period of 06:42:00--06:42:32 UT. We have tested the background coherence levels with the chosen window size (8 s) using random white-noise data and the maximum coherence of the random time series is below 0.6. Thus, the coherence levels are significant for $r>$0.6. KAWs have relatively high coherence up to $\sim$5 Hz at this interval and the phase relation indicates travelling Alfv{\'e}n waves.}
\vspace*{0cm}
\label{figS3}
\end{figure}
\clearpage
\newpage

\begin{figure}
\centering
\vspace*{0cm}\includegraphics[scale=0.6]{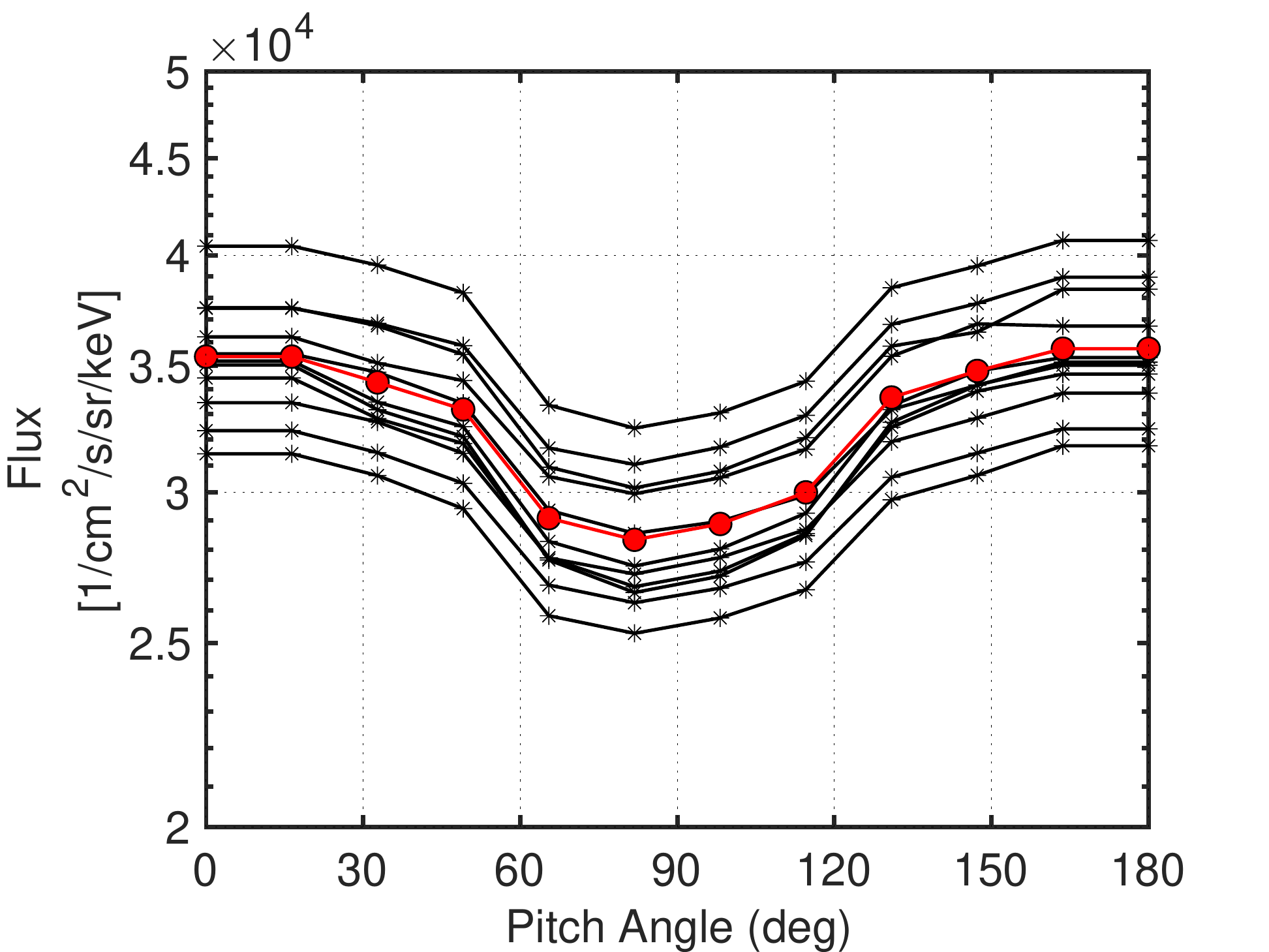}
\caption{MMS1 electron flux pitch-angle distributions measured by FEEPS for energies of 40-200 keV during the period of 06:40--06:50 UT. The red curve is the average distribution. Electron distributions are obtained every $\sim$20 s in spin resolution. Note that MMS can measure electrons outside the loss cone and are unable to resolve the pitch angle within the loss cone (i.e., $<\sim$2$^\circ$).}
\vspace*{0cm}
\label{figS4}
\end{figure}
\clearpage
\newpage

\begin{figure}
\centering
\vspace*{0cm}\includegraphics[scale=0.7]{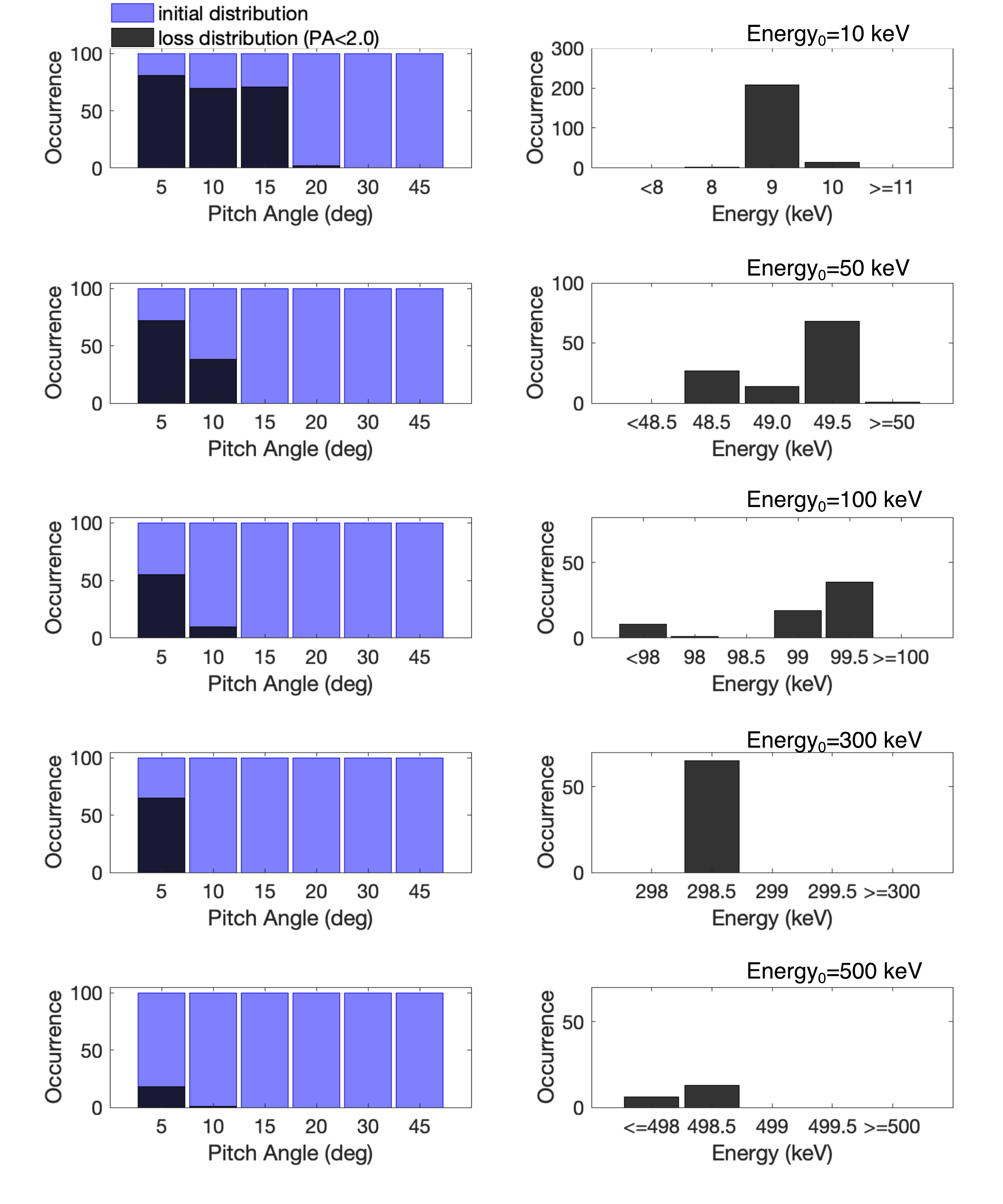}
\end{figure}
\begin{figure}
\caption{(Left panels) Initial (blue) and lost (black) electron pitch-angle distributions for five initial energies of 10 keV, 50 keV, 100 keV, 300 keV, and 500 keV; (Right panels) Electron energy distributions recorded when they are pushed into the loss cone ($<$2$^\circ$) by kinetic Alfv{\'e}n waves. In total 600 electrons are specified with initial pitch angles uniformly distributed at 6 values within 45$^\circ$. This initial pitch-angle distributions have been observed by MMS during the injection. Electron loss rates are calculated by counting the percentage of lost electrons with respect to their initial distributions within 20$^\circ$.}
\vspace*{0cm}
\label{figS5}
\end{figure}
\clearpage
\newpage


%
%
%
%
%
%
%
%
%